\documentclass[sigconf]{acmart}

\usepackage{url}
\usepackage{xcolor}
\definecolor{newcolor}{rgb}{.8,.349,.1}
\usepackage{siunitx}

\definecolor{darkred}{rgb}{0.6,0,0}
\definecolor{lightgreen}{rgb}{0.1,0.8,0.1}
\definecolor{darkgreen}{rgb}{0.1,0.3,0.1}
\definecolor{middlegreen}{rgb}{0.1,0.5,0.1}
\definecolor{darkorange}{rgb}{0.6,0.4,0.2}
\definecolor{darkPurple}{rgb}{0.4,0.1,0.4}
\definecolor{lightblue}{rgb}{0.5,0.5,0.75}
\definecolor{blue}{rgb}{0,0,0.75}
\definecolor{orchid}{rgb}{0.5,0,0.5}
\definecolor{orange}{rgb}{1.0,0.3,0}
\definecolor{darkorange}{rgb}{0.6,0.2,0}

\newcommand{\commentnew}[2]{\textcolor{#1}{#2}}

\newcommand{\yf}[1]{\commentnew{black}{  #1}}
\newcommand{\mn}[1]{\commentnew{black}{ #1}}

\AtBeginDocument{%
  \providecommand\BibTeX{{%
    \normalfont B\kern-0.5em{\scshape i\kern-0.25em b}\kern-0.8em\TeX}}}

\setcopyright{acmcopyright}

\copyrightyear{2020}
\acmYear{2020}
\setcopyright{acmcopyright}\acmConference[IVA '20]{IVA '20: Proceedings of the 20th ACM International Conference on Intelligent Virtual Agents}{October 19--23, 2020}{Virtual Event, Scotland Uk}
\acmBooktitle{IVA '20: Proceedings of the 20th ACM International Conference on Intelligent Virtual Agents (IVA '20), October 19--23, 2020, Virtual Event, Scotland Uk}
\acmPrice{15.00}
\acmDOI{10.1145/3383652.3423882}
\acmISBN{978-1-4503-7586-3/20/09}

\begin{document}

\title{Understanding the Predictability of Gesture Parameters from Speech and their Perceptual Importance}


%

\author{Ylva Ferstl}
\affiliation{%
  \institution{Trinity College Dublin}
  \streetaddress{College Green, Dublin 2}
  \city{Dublin}
  \country{Ireland}
}
\email{yferstl@tcd.ie}

\author{Michael Neff}
\affiliation{%
  \institution{University of California Davis}
  \streetaddress{1 Shields Ave}
  \city{Davis}
  \state{California}}
\email{mpneff@ucdavis.edu}

\author{Rachel McDonnell}
\affiliation{%
  \institution{Trinity College Dublin}
  \streetaddress{College Green, Dublin 2}
  \city{Dublin}
  \country{Ireland}
}
\email{ramcdonn@tcd.ie}

\begin{abstract}

Gesture behavior is a natural part of human conversation. Much work has focused on removing the need for tedious hand-animation to create embodied conversational agents by designing speech-driven gesture generators.
However, these generators often work in a black-box manner, assuming a general relationship between input speech and output motion. As their success remains limited, we investigate in more detail how speech may relate to different aspects of gesture motion. We determine a number of parameters characterizing gesture, such as speed and gesture size, and explore their relationship to the speech signal in a two-fold manner. First, we train multiple recurrent networks to predict the gesture parameters from speech to understand how well gesture attributes can be modeled from speech alone. 
We find that gesture parameters can be partially predicted from speech, 
and some parameters, such as path length, being predicted more accurately than others, like velocity. 
Second, we design a perceptual study to assess the importance of each gesture parameter for producing motion that people perceive as appropriate for the speech. Results show that a degradation in any parameter was viewed negatively, but some changes, such as hand shape, are more impactful than others.
A video summarization can be found at \url{https://youtu.be/aw6-_5kmLjY}.

\end{abstract}




\maketitle

\section{Introduction}

Generating gesture behavior for virtual agents is an important part of making them increasingly life-like and engaging. Much work has focused on generating gestures from speech, but one major challenge has been the large variability of gesture, with gesture choice and expression varying both between speakers as well as within speaker. The same utterance may be accompanied by two completely different gestures even when repeated by the same speaker at different points in time. 
Rather than speech directly informing the gestures to be produced, the \textit{Growth Point} theory of \cite{mcneill1992hand} argues that speech and gesture are two communicative channels both arising from the same cognitive process. Therefore, speech may give us an indication of the underlying intention that inspired a gesture, but may never fully predict the gesture expression. Often however, we want to rely solely on the speech \yf{audio} signal for generating gesture behavior due the ease of obtaining such speech in real applications. \yf{While some other works have included text transcriptions of the speech input \cite{kucherenko2020gesticulator,Ishi2018}, transcription of spontaneous speech can be difficult, and additional processing such as semantic feature extraction may be necessary in particular for smaller datasets.}
For generating gestures from an audio signal, we are interested to what extent we \textit{can} predict the expressive qualities of gesture from speech; specifically, which characteristics of gesture correlate well with the speech signal and can be predicted successfully, and which characteristics are perceptually important.

To this aim, we first ran an exploratory study to investigate how well gesture characteristics may be predicted from a speech signal. We determined a number of gesture parameters, such as speed and range, that describe the expressiveness of a gesture. We then train multiple recurrent networks to model the speech to gesture parameter relationship and discuss their performance.

Secondly, we assessed the perceptual relevance of the gesture parameters in an empirical study. Assessing the perceptual salience of attributes of gesture motion provides guidance on what features must be accurately modeled to produce satisfying animation.

Our focus is on the relationship between speech and the expressive quality of the gestures, so in all cases we maintain the same gesture form as used in the original utterance.
Results indicate that all gesture parameters are predicted above chance, but there is variance in how well they are predicted.  For example, 
arm swivel is predicted better than gesture velocity.
Observers were sensitive to all variations in parameters away from the original performance and increased hand opening was viewed particularly negatively, among other results.  








\section{Related work}

A variety of different approaches have been proposed for the problem of gesture generation from speech. 
Early work employed explicit rule systems mapping text to gestures \cite{Cassell2001,Thiebaux2008,Marsella2013} and statistical modelling of speech features co-occurring with motion features \cite{Neff2008,Bergmann2009}. 
With the rise of machine learning, numerous network types have been investigated, including variations of hidden Markov models \cite{Levine2009,Bozkurt2016}, conditional random fields \cite{Levine2010,Chiu2014}, and restricted Boltzmann machines \cite{Chiu2011}. 
In recent work, recurrent neural networks have proven popular; a classic training loss has been employed for English  \cite{kucherenko2020gesticulator,ferstl2018investigating} and Japanese speech-to-gesture generation \cite{Hasegawa2018,kucherenko2019analyzing}. To combat the problem of mean pose regression in a standard training paradigm, an adversarial training paradigm has been proposed in \cite{ferstl2020adversarial} (similarly for a convolutional network setup in \cite{ginosar2019learning}), and recently, probabilistic generative modelling has shown promise \cite{alexanderson2020style}.
However, due to the highly indeterministic input-to-output relation, modelling plausible gestures remains a difficult problem. There are a multitude of possible gestures for each utterance, and therefore modelling gestures as sequences of joint positions or angles can fail to capture the natural variety of gesture motion. 
We therefore explore alternative representations of gesture that do not rely on explicit joint positions or angles.


Several works have looked at parameter representations of gestures. \cite{zhao2001synthesis} uses Laban Movement Analysis, specifically the Effort and Shape parameters, to describe and modify gestures. \cite{hartmann2005implementing} uses a review of social psychology literature in combination with a gesture corpus analysis to determine a set of six parameters to capture the expressivity of gestures, including gesture scale and fluidity. They find evidence that matching parameters to the communicative intent makes the gesture behavior more appealing.
\cite{Neff2010} uses a similar set of parameters including gesture rate, scale, and position, and find they can significantly influence perceptions of extraversion by modifying these parameters in gestures. \cite{Smith2017} extends this work by using a set of parameter modifications to target perceptions of all Big Five personality traits. 
\yf{\cite{castillo2019we} defines a set of 11 motion parameters and shows that they can manipulate the perceived emotional content, defined by valence and arousal, of a gesture.}

This previous work on parameter representation of gestures shows that we can reliably influence perceptions of personality \yf{and emotion} by applying simple modifications, and gives some evidence that matching measures of gesture expressivity to speech can increase appeal.
While tackling the speech-to-gesture problem, we are interested in which gesture parameters are related to the speech expression. On the one hand, we would like to know which gesture parameters can be successfully predicted from speech. On the other hand, we want to understand which of these parameters are important for perceptually plausible gesture synthesis. 



\section{Dataset \& processing}\label{dataset}

We use a corpus of 6 hours of conversational data, presented in \cite{ferstl2019multi} as our first dataset (dataset A). The dataset consists of high-quality audio and motion recordings of a single right-handed male English speaker producing spontaneous, colloquial speech, in monologue style. 
During network training (Sec. \ref{prediction}), we include dataset B, the open-source Trinity Speech-Gesture dataset \cite{ferstl2018investigating}, a similar corpus of 4 hours of speech and motion data of a different male English speaker (also right-handed). We find that including this dataset improves performance. 
We segment the gesture databases using the stroke phase labels (see \cite{ferstl2020adversarial}).

\subsection{Speech processing}\label{speechProcessing}
We tested the suitability of three different feature sets for speech processing. The first set consists of the 12 Mel-frequency cepstral coefficients (MFCCs), common in speech recognition 
as well as previous speech-gesture work \cite{kucherenko2019analyzing, ferstl2020adversarial}. Secondly, we tested Geneva Minimalistic Acoustic Parameter Set (GeMAPS), both the 18 features of the compact version, as well as an extended set of 23 features presented in \cite{Eyben2016}. The GeMAPS has been specifically developed for affect recognition.
Finally, we tested a three feature set simply consisting of the pitch (F0), plus its first and second derivative to describe change over time.
We extracted all speech features using OpenSMILE \cite{eyben2013recent}.
After training a number of speech-to-gesture-parameter models in an exploratory manner with each of the three feature sets, we found GeMAPS to work best overall, as measured by the numeric loss during training, with the compact and the extended feature set performing similarly.
MFCCs performed well but slightly worse than GeMAPS, and the feature set of pitch plus derivatives greatly underperformed. We will therefore report results using the GeMAPS input representation. 

\subsection{Gesture processing}
We aimed to find a number of gesture characteristics that could describe the expression of a gesture. We define these characteristics based on the central part of a gesture, the stroke phase, which represents the expressive phase of a gesture and carries its meaning \cite{mcneill1992hand} \yf{or, in the case of non-meaningful beat gestures, represents the period of the highest effort}. The stroke phase was determined following a previous approach by \cite{ferstl2020adversarial}, using the hand-annotation where available, and the automatic stroke classification otherwise. 
\mn{Each feature below is calculated for each gesture from the corresponding motion capture data.}
\begin{enumerate}
    \item velocity
    \item initial acceleration
    \item gesture size:
    \subitem (3.1) path length
    \subitem (3.2) major axis length
    \item arm swivel
    \item hand opening
\end{enumerate}



Velocity and initial acceleration both describe the kinematics of the gesture, represented by the maximum stroke velocity (1), and by the mean acceleration to the first major velocity peak (2). Velocity captures a character's tempo and relates to the amount of energy they are using. Initial acceleration may be useful to model an emphatic gesture start.  This is akin to the type of tangent adjustment done between key frames in hand animation. 

With gesture size (3), we describe the spatial extent of the gesture. We measure this in two ways: The total path length of the gesture stroke, \yf{calculated by summing the difference between the wrist positions at each subsequent frame}, and the length between the minimum and maximum point of the stroke, which we will subsequently refer to as major axis length. 

\mn{Arm swivel (4) describes the rotation around an axis between the shoulder and the wrist, bringing the elbow in or away from the body.} 
This angle modifies the amount of space taken up by the gesture and can change the perceived personality \cite{Smith2017} and has been postulated to relate to humility and arrogance~\cite{shawn1963every}.

The last parameter, (5), describes the hand shape during a gesture, specifically, how open or closed the hand is. We calculate this as the mean distance of the finger tips (excluding thumb) from the base of the wrist. Such variation in hand flexion has been shown to impact the perception of character personality \cite{wang2016assessing}.

Based on previous work, we expect gesture velocity and acceleration to be well predicted from speech (e.g. \cite{pouw2019gesture,loehr2012temporal}), whereas more uncertainty around speech correspondence to arm swivel and hand opening.




\section{Gesture parameter prediction}\label{prediction}

\begin{figure}
  \begin{minipage}[c]{0.15\textwidth}
    \includegraphics[width=\textwidth]{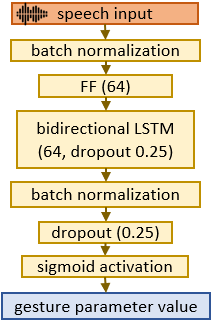}
  \end{minipage}
  \begin{minipage}[c]{0.33\textwidth}
    \caption{
       Network structure of the speech-to-gesture-parameter models. Speech input is batch-normalized, then passed through a linear feed-forward layer (FF) of size 64. The core of the model is a bidirectional LSTM cell of size 64 (25\% input dropout). The output of the recurrent cell is batch-normalized and 25\% dropout is applied before the final output layer with sigmoid activation.} 
       \label{network}
  \end{minipage}
\end{figure}

The first part of our work focuses on the problem of predicting gesture characteristics from a speech signal. Our aim hereby is to assess which gesture descriptors can be predicted from speech with current machine learning techniques. \yf{While it is reasonable to consider using a single model to jointly predict all gesture properties, we found this in practice difficult to optimize and instead model one gesture property at a time.}
We tested a number of model configurations, such as one vs. two network layers, as well as testing different layer sizes. Our goal was to adopt a fairly standard network architecture to explore if gesture parameters can be predicted from speech in such a framework. After experimenting with configurations, we found the general network structure in Fig. ~\ref{network} performed best. Using more layers or larger layer sizes led to frequent over-fitting; using smaller layer sizes or simpler layers (uni-directional instead of bidirectional recurrent layer) led to under-fitting. 

We use recurrent neural networks due to their strength in modelling sequential time-series data as well as their use in recent speech-to-gesture research \cite{Hasegawa2018, kucherenko2019analyzing,ferstl2018investigating,ferstl2020adversarial}.
All models take an input sequence of speech features, extracted over the period of the corresponding gesture's stroke phase plus a context of 1 second in each direction. Sequence-based models require a constant input length \yf{within a training batch}, we therefore define a maximum input length of 5.5 seconds, based on the maximum stroke duration found in the datasets plus context windows. All shorter sequences are zero-padded to fulfill the constant input length requirement.

The model applies batch normalization to the input, then input transformation through a feed-forward layer. This is followed by one recurrent network layer, followed by batch normalization and a dropout layer for regularization purposes.  
The outputs of a model are the values of the gesture parameter under investigation (e.g., velocity, initial acceleration, etc.), normalized to the range of 0-1 for each given stroke, one value for each hand.  The output nodes have a sigmoid activation. Training minimized the mean squared error between predicted and true value. 
\yf{We use the Adam optimizer with a learning rate of $2\times10^{-4}$ and standard decay.}

To generate output, it is necessary to resolve a potential ambiguity between the predicted behavior of each hand.
The stroke label does not include the handedness, i.e. whether the right, the left, or both hands are performing a stroke. Therefore, the predictive model must make some assumptions about the active hand(s). 
The model can learn general statistics regarding differences between the two hands (e.g. left hand generally slower), but will not be able to predict diverging values indicating gesture handedness (e.g. high velocity for right hand and zero velocity for left hand, indicating a right-handed gesture)\yf{, unless successfully inferring handedness from the audio signal. Labelling handedness could improve future modelling approaches.}  


\subsection{Results}\label{predResults}

\begin{table*}
\begin{center}
      \caption{Performance evaluation of the speech-to-gesture-parameter models. In brackets are random sampling errors, computed using a randomly drawn parameter sample from true values as prediction value. \yf{Random samples for all parameters except gesture size (path length and major axis length) are drawn with the constraint noted in Eq. \ref{pathLenRestr}. $MAD$ denotes the mean absolute deviation of each gesture parameter. We report mean (\large{$\bar{e}$})\normalsize  and median (\large{$\tilde{e}$}\normalsize) errors for the left (L) and right (R) hand, as well as the \% reduction (red.) in error between random sampling and our models.  A higher reduction percentage implies better model performance.}}
      \label{predictionResultsTable}
      \begin{tabular}{l*{10}{c}r}
        \toprule
        & MAD L/R & \LARGE$\bar{e}$ \normalsize L & red. & \LARGE$\bar{e}$ \normalsize R & red. & \LARGE$\tilde{e}$ \normalsize L & red. & \LARGE$\tilde{e}$ \normalsize R & red. \\
        \midrule
         velocity ($m/s$) & 0.34/ 0.37 & 0.31 (0.38) & 19\% & 0.35 (0.43) & 17\% & 0.24 (0.27) & 12\% & 0.28 (0.31) & 11\%\\
         initial acceleration ($m/s^2$) & 0.30/ 0.34 & 0.27 (0.37) & 27\% & 0.30 (0.42) & 28\% & 0.15 (0.18) & 17\% & 0.16 (0.23) & 30\% \\
         path length ($m$) & 0.21/ 0.24 & 0.11 (0.28) & 60\% & 0.13 (0.33) & 61\% & 0.06 (0.17) & 63\% & 0.07 (0.21) & 64\%\\
         major axis length ($m$) & 0.10/ 0.11 & 0.08 (0.14) & 41\% & 0.09 (0.16) & 45\% & 0.06 (0.10) & 39\% & 0.06 (0.12) & 49\%\\
         arm swivel (\textit{degrees}) & 12.42/ 10.16 & 11.52 (15.39) & 25\% & 9.32 (13.28) & 30\% & 8.82 (10.47) & 18\% & 6.86 (10.02) & 31\%\\
         hand opening ($cm$) & 3.91/ 3.73 & 1.64 (2.29) & 29\% & 1.23 (1.85) & 33\% & 1.14 (1.39) & 18\% & 0.97 (1.19) & 18\%\\
      \bottomrule
    \end{tabular}
    \end{center}
\end{table*}
Using the stroke phases as our segmentation, our training data consists of a total of almost 23,700 gesture stroke samples, with approximately 58\% stemming from dataset A \cite{ferstl2020adversarial} and 42\% from the dataset B \cite{ferstl2018investigating}. We hold back about \yf{4\% of the samples for validation and 1.5\% for testing}, chosen randomly. The velocity and acceleration models reached best performance after 70 epochs, all other models were trained for about 140 epochs.
As described in Sec. \ref{speechProcessing}, we found GeMAPS input speech representation to work best overall,
the reported models used the compact set in the case of gesture size, and the extended set in all other cases.

\mn{Our goal is to understand which gesture features are predictable from  speech audio.  A straightforward performance measure is the mean error between predicted and actual parameter values across all gestures. As the mean can be distorted by small numbers of large errors, we also report the median error. Taken alone, this does not say if the audio data is informative. We therefore compare the predictions to a model that has no audio input. In this case, all we can do is match the underlying statistics, the mean and variance, of our gesture data for each parameter. We do this by randomly selecting gestures for each slot and report the error such an approach would produce. Comparing these two errors gives an indication of how much the audio improved predictions. These errors are reported in Table \ref{predictionResultsTable}, with errors for random selection in brackets. We drew the samples in a database-specific manner, i.e. we always use the correct database to draw from for each sample, to ensure that simple speaker-detection by the model is not the reason for superior performance. We compute the random sampling error three times for each gesture parameter and report the average values.} 
\yf{Secondly, we report the standard deviation (std) as well as the mean absolute deviation (MAD) for all true parameter values; a model with no prediction power can be expected to yield an error similar to MAD.}
\yf{We trained all networks n times, reporting average error and std. As we found std to be low in all cases, we considered n=3 to be sufficient. Prediction results and random as well as MAD baseline values are listed in Table \ref{predictionResultsTable}.}

\textbf{Maximum velocity} averaged 0.59 $m/s$ (std=0.46 $m/s$, MAD= 0.34 $m/s$) for the left, and 0.70 $m/s$ (std=0.49 $m/s$, MAD=0.37 $m/s$)) for the right hand, and our model produced mean errors of 0.31 $m/s$ (std=0.01 $m/s$) and 0.35 $m/s$ (std=0.01 $m/s$) respectively, with the median at 0.24 $m/s$ and 0.28 $m/s$. Referring to Table \ref{predictionResultsTable}, we see 19\% and 17\% mean error reduction for the left and right hand, compared to random sampling, and 12\%  and 11\% median error reduction.
The model avoids very low velocity predictions, and to some degree high velocity predictions (Supp. Mat. Fig. 1 (top)). \yf{A possible reason is the use of the mean squared error function in training, which can encourage outputs to stay around the mean value.}


\textbf{Initial acceleration} averaged  0.16 $m/s^2$ (std=0.58 $m/s^2$, MAD=0.30 $m/s^2$) for the left, and 0.22 $m/s^2$ (std=0.61 $m/s^2$, MAD=0.34 $m/s^2$) for the right hand, and our model produced mean errors of 0.27 $m/s^2$ (std=0.01 $m/s^2$, median=0.15 $m/s^2$) and 0.30 $m/s^2$ (std=0.01 $m/s^2$, median=0.16 $m/s^2$), respectively. The model again avoids very high acceleration predictions, however, high acceleration is often correctly identified though the predicted value tends to be lower than the true value (see plotted prediction results in Supp. Mat. Fig. 1 (bottom)). Compared to our baseline random sampling error, we achieve a mean error reduction of 27\% and 28\% for the left and right hand, respectively, and 17\% and 30\% median error reduction (see Table \ref{predictionResultsTable}).
\yf{The mean error reductions indicate that acceleration can be modelled more successfully than velocity.}




Our first measure of \textbf{Gesture Size} is \textit{path length}. We find that gesture path length is highly correlated with the length of the corresponding input speech segment; a longer speech input is associated with a longer stroke. Hence, in addition to comparing prediction results to the random sampling error (see Table \ref{predictionResultsTable}), we employ a second test taking into account only speech length. For this, model input is a single speech feature has the value 1 for all input time steps before the zero-padding. This input processing means that the model can base predictions solely on the length of the input signal, without receiving information about the speech quality. 
Using only speech length versus GeMAPS input yielded very similar errors. Mean path lengths were 0.25 $m$ (std=0.30 $m$, MAD=0.21 $m$) and 0.32 $m$ (std=0.34 $m$, MAD=0.24 $m$) for the left and right hand, respectively. Using only speech length input yielded mean errors of 0.11 $m$ (std=0.00 $m$, median=0.07 $m$) and 0.13 $m$ (std=0.00 $m$, median=0.08 $m$) for the left and right hand, while using GeMAPS results in mean errors of 0.11 $m$ (std=0.00 $m$, median=0.06 $m$) and 0.13 $m$ (std=0.00 $m$
median=0.07 $m$), respectively.
\yf{Paired Wilcoxon tests showed no significant improvement of path length prediction for GeMAPS input over speech length input, suggesting that the length of the speech signal was the essential determinant for path length.}




Our second measure of \textbf{Gesture Size}, is the \textit{major axis length}, defined as the length of the axis between the minimum and maximum point of the gesture. The average major axis lengths for the left and right hand are 0.15 $m$ (std=0.14 $m$, MAD=0.10 $m$) and 0.19 $m$ (std=0.15 $m$, MAD=0.11 $m$), respectively, and our model produced mean errors of 0.08 $m$ (std=0.00 $m$, median=0.06 $m$) and 0.09 $m$ (std=0.00 $m$,median=0.06 $m$) for the left and right hand, respectively (see also Supp. Mat. Fig. 2 (bottom)). We critically evaluate the results for the major axis length in the same manner as for the path length, using only speech length as input. 
Model errors the same as for GeMAPS input, and paired Wilcoxon test showed speech input to yield no significantly better performance. 
\yf{As for path length predictions, this suggests that major axis length predictions were only significantly informed by the length of the speech signal.}

\yf{Due to the strong correlation of speech input length and gesture size, we tighten the conditions for the random baseline sample selection. In addition to drawing samples dataset-specific, we also restrict sample selection to a small range around the true gesture size. That is, when selecting a random sample, we only consider samples $i$ for which the path length $pl$:}
\begin{equation}\label{pathLenRestr}
    pl_{true}-\frac{std(pl)}{4} < pl_{i} < pl_{true}+\frac{std(pl)}{4}
\end{equation}
This represents a tight restriction of random selection to only around 5\% of the total samples. All random sampling results reported are path length restricted.

For \textbf{arm swivel}, increasing swivel angle for the left arm (moving the elbow out) means a higher positive value, whereas increasing the right arm's swivel means increasingly negative values. The left arm had a mean angle of \ang{14.43}  (std=\ang{16.28}, MAD=\ang{12.42}), and our model yielded a mean error of \ang{11.52} (std=\ang{0.34}, median=\ang{8.82}) . 
The mean right swivel was \ang{-21.58} (std=\ang{13.61}, MAD\ang{=10.16}) and our model yielded a mean error of \ang{9.32} (std=\ang{0.05}, median=\ang{6.86}) (Supp. Mat. Fig. 3 (top)). Mean error reductions with respect to random sampling were 25\% (left hand) and 30\% (right hand), and median reductions were 18\% (left hand) and 31\% (right hand) (see Table \ref{predictionResultsTable}).


\textbf{Hand opening} averaged 16.56 $cm$ (std=4.37, MAD=3.91) and 17.05 $cm$ (std=4.10, MAD=3.73) for the left and right hand, respectively, and corresponding mean model errors were 1.64 (std=0.00, median=1.14) and 1.23 (std=0.02, median=0.97) (see also Supp. Mat. Fig. 3 (bottom)). As noted in Table \ref{predictionResultsTable}, this meant mean error reductions, with respect to random sampling, of 29\% and 33\% for the left and right hand, respectively, and median reductions of 18\% each.  


Shapiro-Wilk tests showed that the distributions of gesture parameters were not normal. We evaluated error reduction with respect to random sampling using paired Wilcoxon tests with Bonferroni correction for multiple hypothesis testing (n=16). All models performed better than random sampling (all p<.001/16).

Wilcoxon tests further revealed that path length prediction errors were lower than for all other parameters except right arm swivel (all p<.001). Arm swivel errors were lower compared to all parameters except path length and left hand acceleration (p<.001/16 for all but left major axis length (p<.05/16)). 

\subsection{Discussion}
In this first part of our work, we sought to examine which gesture parameters may be predicted well from a speech signal and may therefore be well accounted for by a speech-to-gesture generation model. 
For this, we explored five different gesture parameters.

Gesture velocity has been used in previous work on gesture generation from speech \cite{Hasegawa2018,kucherenko2020gesticulator}. However, interestingly, we found this to be a difficult parameter to model from speech. While there does appear to be an underlying relationship between velocity and the speech representaion, it proved to be difficult to capture velocities farther from the mean, i.e. we could not capture the full variability of velocities.
As an additional measure of gesture kinematics, we modelled the acceleration to the first major velocity peak. Acceleration was predicted more accurately than velocity. The model often successfully detects high initial acceleration; common errors are failing to capture high initial acceleration of the left hand (non-dominant hand) and instead only capturing this for the right hand, as well as not modelling very high values.
Avoidance of high value predictions can be expected due to the low frequency of these values overall; the model would be penalized strongly for wrongly predicting large values, and rewarded only in the infrequent cases of true high values. Oversampling high values to increase their frequency could help encourage more diverse predictions.

For modelling gesture size, we used two measures, path length and major axis length.
We found that the gesture size measures were predicted best overall, however
as larger lengths may take more time to complete, we compare our model to a baseline prediction model conditioned on only the length of the speech signal. The length of the speech signal was highly correlated with gesture path length and major axis length, and statistical tests showed using speech input did not improve predictions.
Our results also emphasize the difficulty of the speech-to-gesture generation problem. Even with a highly reduced data complexity of just one gesture descriptor rather than many skeleton joints, modelling remains difficult.
While motion parameter predictions based on audio showed lower error compared to baselines, indicating that audio is informative when determining gesture parameters, the errors in these predictions are still relatively large when compared against expected deviations for these parameters. This may suggest that audio alone is not sufficient for predicting gesture parameters.


\section{Gesture parameter evaluation}\label{perception}


As audio is only partially successful at predicting gesture parameters, we want to understand which gesture parameters must be accurately realized in order to achieve satisfying motion.  To explore this, we design an empirical evaluation of the impact of gesture parameters on perception. We assess people's judgment of the gesture expression regarding its suitability for the expressed speech. 
We test the perceptual impact of our gesture parameters by creating variations that increase or decrease them, as described below. 

\subsection{Stimuli creation}\label{stimuli}
Artificial stimuli are created through a three step process.  First, the variation in the source data is measured.  Second, clips are selected that best represent high and low variations within this. Third, these clips are algorithmically modified to fully match the desired high and low performance.

First, we compute the natural variation of each of our parameters within the gesture database by calculating the 25th percentile marker as a lower bound, and the 75th percentile marker as the upper bound. 
Samples below the lower bound are defined as having a \textit{low} expression, and samples above the upper bound are defined as having a \textit{high} expression of a given parameter. \yf{(The data distribution is visualized in Fig. 4 in Supp. Mat.)}.

Second, we randomly select short gesture sequences of about 10 seconds. A 10-second time-frame has previously been shown to be sufficient for participants to make judgements about conversing agents \cite{Ennis2010}. 
For \textit{low} sequences, we use sequences that contain \textit{low} parameter expressions. However, as there are practically no 10 second sections in the database of only \textit{low} expression, we allow the sequences to contain \textit{medium} expression (values below the upper bound), but give preference to gesture sequences with the highest percentage of \textit{low} samples. 
Equivalently, for \textit{high} sequences, we use sequences containing mainly \textit{high} expression, allowing some \textit{medium} expression samples. This biased selection ensures that the edited clips are as different as possible from the source clips (i.e. error maximizing). 

In the third step, we create the parameter manipulations. For \textit{low} sequences, we increase the parameter expression to \textit{high}, keeping within the found natural limits. For \textit{high} sequences, we decrease the parameter expression to \textit{low}. We select 5 samples each for the \textit{low} and the \textit{high} manipulations of each parameter.
As baseline samples, we randomly select 10 sequences that remain un-manipulated.

All samples are generated with animation software \yf{based on the open-source animation environment DANCE \cite{shapiro2005dynamic}} that uses a motion parameterization similar to \citet{neff2009interactive} and IK tools to generate variations of the input motion capture data.
It takes as input the motion data and the corresponding stroke labels and synthesizes preparation (bringing the hands into position for the gesture) and retraction (returning the hands to a rest position) phases for the strokes \mn{using splines}, proportionally matching the stroke speed. \mn{Synthesizing preparations and retractions avoids problems such as two lengthened gestures not maintaining the necessary time for a retraction that was originally present between them.} 
If a manipulation is applied, it is applied to the stroke phase. 
We restrict our data selection to the hand-annotated sections of dataset A . Including dataset B in this step would require manually correcting all automatically determined stroke labels to ensure correct boundaries.
All stimuli can be viewed at \url{https://www.youtube.com/playlist?list=PLLrShDUC_FZzhemzr0g1ekt1jz45-y_u3}.

\subsection{Experiment }



The experiment was designed with the Unity3D game engine and the \yf{open-source} Virtual Human Toolkit (VHTK) \cite{hartholt2013all}. The displayed character was Brad from the VHTK, 
producing regular eye blinks, lip synchronisation, as well as an idle motion for the body excluding the arms and hands. 
In each experiment trial, participants first watched a ~10 second clip of the character acting out one of the gesture sequences. Following the clip, participants were asked the following question:
\textit{"How well did the expressive quality of the gestures match 
the expressive quality of the speech?"}

This question was specifically designed to motivate participants to focus on the expression of the gestures; we did not want participants to judge the semantic entropy of the gesture sequence.
A 7-point Likert scale was provided as a rating scheme.
Participants first completed 5 example trials for which responses were not recorded. This was in order to establish an expectation of the gesture quality variation in the experiment and to familiarize the participants with the rating scale.
Following the example trials, participants completed 60 experiment trials (5 samples for each of the 5 parameters, with 2 expression manipulations each, plus 10 baseline samples), presented in random order.

The online experiment was distributed via university mailing lists\yf{, with an incentive of a 100 Euro raffle voucher}. We collected data from 60 participants (23 females, 36 males, 1 other gender, ages 18-59 years, \textit{M} = 26.4, \textit{SD} = 9.1), all of whom gave informed consent regarding their participation. All participants reported sufficient English proficiency (35 ``native'', 20 ``fluent'', 3 ``very good'', 2 ``good'').

\subsection{Results}

\begin{figure}[t]
  \centering
  \includegraphics[width=8cm]{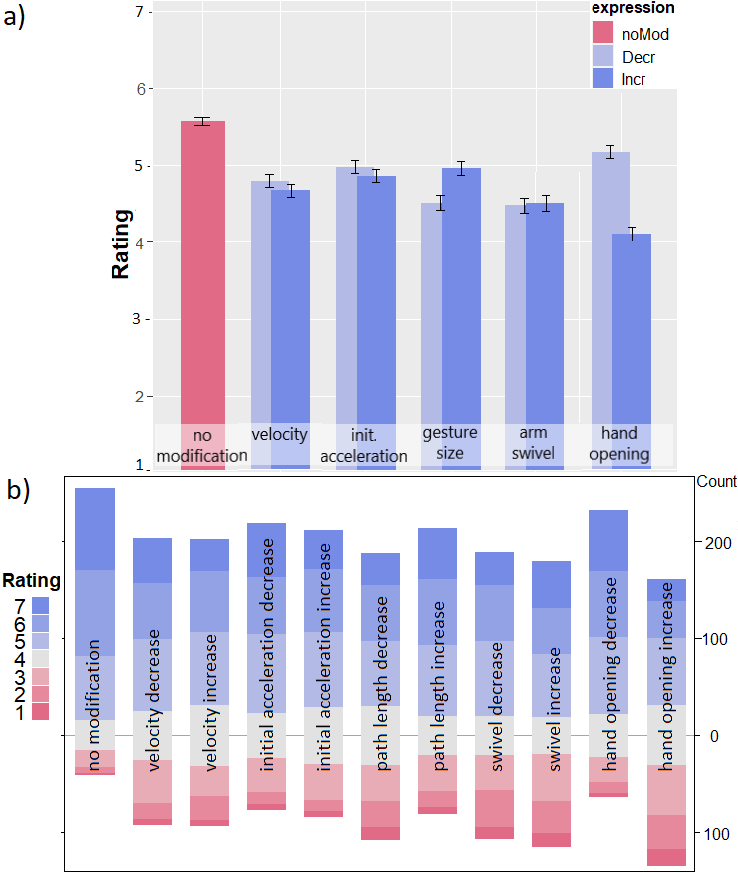}
  \caption{Perceptual results. a) Mean rating scores for all experimental manipulations. Unmodified gestures received the highest average rating, and increased hand opening the lowest. 
  b) Stacked bar chart of all given ratings. Plotted is the frequency of responses for the 7 rating scores. (The y-axis represents the frequency of responses). The no-modification condition is scaled by 50\%.}
  \Description{}
  \label{perceptualResults}
\end{figure}

The study consisted of two factors, the {\em parameter} that was modified and the {\em direction} of the modification. The first factor had 11 conditions, with mean ratings summarized in Fig. \ref{perceptualResults} a), and the rating score distribution further explored in Fig. \ref{perceptualResults} b). The second factor had two levels, increase and decrease.



We analyzed the data by treating the rating scores as ordinal data and fitting a cumulative link model, using clm from the R ordinal package \cite{ordinal}. 
All modification conditions were rated significantly lower than the no modification condition (all p<0.001, Bonferroni corrected with n=55).
Decreasing gesture size was rated significantly worse than increasing (p<.05).
Decreasing hand opening was preferred over increasing (p<.001). Increasing hand opening received the lowest rating compared to all other conditions (all p<.05).
Complete results are detailed in Table \ref{perceptualResultsTable}.


\begin{table*}
      \caption{All results for the perceptual experiment. Indicated are both significant and non-significant condition differences (plotted in Fig. \ref{perceptualResults}). + means the row condition was rated higher, - means lower rating. \yf{All p values were Bonferroni corrected (n=55): , $*=p$<.05/55, $**=p$<.01/55, $***=p$<.001/55. n.s.=not significant}}
      \label{perceptualResultsTable}
      \begin{tabular}{l*{11}{c}r}
        \toprule
         & no mod. & velocity $\downarrow$
 & velocity $\uparrow$ & init. acc. $\downarrow$
 & init. acc. $\uparrow$ & size $\downarrow$
 & size $\uparrow$ & swivel $\downarrow$
 & swivel $\uparrow$ & hand $\downarrow$
 &hand $\uparrow$ \\
        \midrule
        no mod. & - & $***^{+}$ & $***^{+}$ & $***^{+}$ & $***^{+}$ & $***^{+}$ & $***^{+}$ & $***^{+}$ & $***^{+}$ & $**^{+}$ & $***^{+}$  \\
        velocity $\downarrow$ & $***^{-}$  & - & n.s. & n.s. & n.s. & $*^{+}$ & n.s. & n.s. & n.s. & $*^{-}$ & $***^{+}$ \\
        velocity $\uparrow$ & $***^{-}$  & n.s. & - & n.s & n.s. & n.s & n.s. & n.s.& n.s. & $**^{-}$ & $**^{+}$ \\
        init. acc. $\downarrow$ & $***^{-}$ & n.s. & $*^{+}$ & - & n.s. & $*^{+}$ & n.s. & $*^{+}$ & $*^{+}$ & n.s. & $***^{+}$ \\
        init. acc. $\uparrow$ & $***^{-}$ & n.s. & n.s. & n.s. & - & n.s. & n.s. & n.s. & n.s. & n.s. & $***^{+}$\\
        size $\downarrow$  & $***^{-}$ & n.s. & n.s. & $*^{-}$ & n.s. & - & $*^{-}$ & n.s. & n.s. & $***^{-}$ & $*^{+}$ \\
        size $\uparrow$  & $***^{-}$ & n.s. & n.s. & n.s. & n.s. & $*^{+}$ & - & $*^{+}$ & $*^{+}$ & n.s. & $***^{+}$\\
        swivel $\downarrow$  & $***^{-}$ & n.s. & n.s. &  n.s. &  n.s. & n.s. & $*^{-}$ & - & n.s. & $***^{-}$ & $*^{+}$\\
        swivel $\uparrow$  & $***^{-}$ & n.s. & n.s. & $*^{-}$ &  n.s. & n.s. & $*^{-}$ & n.s. & - & $***^{-}$ & $*^{+}$\\
        hand open $\downarrow$  & $***^{-}$ & $*^{+}$ & $**^{+}$ & n.s. & n.s. & $***^{+}$ & n.s. & $***^{+}$ & $***^{+}$ & - & $***^{+}$\\
        hand open $\uparrow$ & $***^{-}$ & $*^{-}$ & $*^{-}$ & $***^{-}$ & $***^{-}$ & $*^{-}$ & $***^{-}$ & $*^{-}$ & $*^{-}$ & $***^{-}$ & - \\
      \bottomrule
    \end{tabular}
\end{table*}

\subsection{Discussion}

We found that all our gesture modification had a significant perceptual effect. Unmodified gestures were preferred over all modification conditions, indicating some perceptual relevance for each of the five gesture parameters. 

Altered gesture kinematics, as described by gesture velocity and initial acceleration, significantly worsened speech-gesture match, with the slowing-down modification yielding similar ratings as the sped-up modification. \yf{In Sec. \ref{prediction}, we found gesture velocity particularly difficult to model. The perceptual impact of velocity in our study suggests the need for more work on modelling velocity well.}

For modified gestures, we found that enlarged gesture size was preferred over reduced size gestures. Enlarged gesture size was further preferred over a number of other modifications, while reduced gesture size showed the opposite trend. Machine learning models for gesture generation are often trained with a mean-squared error loss \cite{Hasegawa2018,kucherenko2020gesticulator,ferstl2018investigating}, commonly leading to smaller than natural gestures due to convergence to the mean pose. Our perceptual results give further motivation to move away from such traditional model training approaches. Recent works have proposed alternative approaches \cite{alexanderson2020style,ginosar2019learning,ferstl2020adversarial}. 
\yf{While we achieved good modelling results for gesture size, this was due to a strong correlation of gesture size and input speech length. Therefore, to infer correct gesture size, focus should be on determining the correct window (size) for a gesture.}

There was a large effect of hand opening, with the open, flat hand rated significantly lower than all other modifications. Gesture sequences with decreased hand opening were preferred over most other modifications. Strong effects for manipulating hand shape has also previously been reported by \cite{wang2016assessing} in a study on personality perceptions. 
Modelling finger motion is a complex problem due to the high dimensionality of the hand skeleton; when accurate hand shape prediction is not possible, based on our results, we suggest animating slightly flexed fingers rather than straightened fingers.

Modifying arm swivel angle in either direction elicited relatively low preference ratings, indicating this to be an important factor in believable gesture synthesis. \yf{Notably, arm swivel was also predicted relatively well in Sec. \ref{prediction}.}



\section{General Discussion}

In this work, we investigated the relationship between speech and gesture expressivity. Gesture generation approaches often assume some underlying connection between modalities by training black-box models, feeding in speech data and outputting high-dimensional and complex skeleton motion data. Due to their limited success, we aimed to assess in more detail how speech may relate to gesture motion. Based on a literature review, we first determined a number of parameters to characterize gesture. We then assessed the speech-gesture parameter relationship in two ways.

First, we used machine learning, specifically recurrent neural networks, to phrase the question as a problem of predicting gesture properties from speech. We train separate models for each gesture parameter, working solely on the audio speech signal as input. By judging the successes or failures of the model predictions, we gain a measure of how well the speech signal relates to a given gesture parameter. Results indicate that all gesture parameters are predicted above chance, but there is variance in how well they are predicted. For example, the size of a gesture is predicted better than its velocity. Arm swivel predictions, surprisingly, surpass all other measures but path length. Our results also indicate the remaining difficulty in modelling the speech to gesture relation. Previous work on gesture generation has reported good adherence of their model to the acceleration distribution of a dataset \cite{kucherenko2020gesticulator,kucherenko2019analyzing}, however, our results indicate that the correct acceleration at the correct time matters, and generated gestures should hence be assessed in a gesture-specific rather than output-general manner. \yf{Rather than only assessing acceleration distribution of the output, evaluation should consider the correctness of the acceleration per gesture}. 

\mn{Finally, while gesture parameter predictions were significantly above baseline, they remained well short of ground truth, indicating that audio alone may not be sufficient to predict gesture performance.} \yf{Future work could consider additional input, such as semantic content via information extraction from speech transcripts. Other types of models could also be explored for the task.}
\yf{Our gesture property modelling is limited to two speakers; it is unclear if our results represent `typical' speech-gesture relationships, or if a larger set of speakers would yield different results.}


Second, we conducted a perceptual study to assess the relevance of each gesture parameter for gesture synthesis. For this, we manipulated the level of each parameter and tested the impact on the perceived match of speech and gesture. Observers were sensitive to all variations in parameters away from the original performance, indicating that each of our chosen parameters is important in realistic gesture synthesis. Hand pose showed to be particularly important, with flat, open hands being viewed especially negatively, and more flexed fingers being preferred. Regarding gesture size, we found enlarged gestures being preferred over reduced gestures.

For gesture parameter prediction, we see an expected preference of the models to keep predictions somewhat around the mean for all parameter values, infrequently predicting extreme values. Based on our perceptual results, speech-to-gesture training data could be augmented for better results: for example, due to participants' preference for enlarged versus reduced size gestures, and the common problem of reduced-size gesture output in machine learning models, we could increase the frequency of large gestures within the training dataset. This could be done in three ways: by oversampling large gestures selectively, by oversampling and augmenting large gestures by applying perceptually less salient modifications (e.g. slight acceleration warps), or by applying data augmentation of smaller gestures (artificially enlarging). Additionally, rather than tackling high-dimensional finger motion modelling, simply using slightly flexed fingers is a perceptually reasonable choice. 

With this work, we provide better insights into which aspects of gesture may be modelled from speech. We suggest a step toward better evaluation of gesture generation models by providing numeric gesture descriptors that impact the perceived match of the generated gesture, as shown by our perceptual study.





In future work, we want to address the problem of determining the gesture timing from speech, without relying on motion data.
Our gesture modelling results are limited to two speakers, and our perceptual results to one speaker. In future work, we would like to include a larger variety of speakers and speaker style.
While this work focused on performance variation, it is also important to correctly match the semantics of the gesture with the spoken text.  Systems that generate gesture from speech signals will ultimately need to match both style and content.
As a next step, we would like to explore gesture generation based on parameterization, avoiding the problem of high-dimensional skeleton data.

Supplemental Figures can be found at \url{https://tinyurl.com/y4hdefho}

\begin{acks}
 
This work was funded by Science Foundation Ireland under the ADAPT Centre for Digital Content Technology (Grant 13/RC/2106) and supported by TCHPC (Research IT, Trinity College Dublin). 
\end{acks}


\bibliographystyle{ACM-Reference-Format}
\bibliography{bibliography}

\end{document}